\documentclass[namedreferences]{solarphysics}

\usepackage[optionalrh]{spr-sola-addons} 
\usepackage{graphicx}        
\usepackage{color}           

\newcommand{\eg}{\textit{e.g.}}
\newcommand{\ie}{\textit{i.e.}}

\newcommand{\aap}{    {\it Astron. Astrophys.}}
\newcommand{\aaps}{   {\it Astron. Astrophys. Suppl.}}
\newcommand{\aapr}{   {\it Astron. Astrophys. Rev.}}

\newcommand{\apj}{    {\it Astrophys. J.}}
\newcommand{\apjl}{   {\it Astrophys. J. Lett.}}
\newcommand{\apjs}{   {\it Astrophys. J. Suppl. Ser.}}

\newcommand{\araa}{   {\it Annual Review of Astron. \& Astrophys.}}
\newcommand{\baas}{   {\it Bulletin of the American Astron. Soc.}}

\newcommand{\solphys}{{\it Solar Phys.}}

\newcommand{\ssr}{    {\it Space Sci. Rev.}}
\newcommand{\mdash}{-}
\chardef\us=`\_

\begin{document}

\begin{article}
\begin{opening}

\title{Submillimeter Radiation as the Thermal Component of the Neupert Effect}

\author[addressref={aff1}]{\inits{J.F.}\fnm{Jorge~F.}~\lnm{Valle~Silva}\orcid{0000-0001-7144-7967}}
\author[addressref={aff1,aff2},corref,email={guigue@craam.mackenzie.br}]{\inits{C.G.}\fnm{C.~Guillermo}~\lnm{Gim\'enez~de~Castro}\orcid{0000-0002-8979-3582}}
\author[addressref={aff1,aff3}]{\inits{P. J. A.}\fnm{Paulo~J. A.}~\lnm{Sim\~oes}\orcid{0000-0002-4819-1884}}
\author[addressref={aff1}]{\inits{J.-P.}\fnm{Jean-Pierre}~\lnm{Raulin}\orcid{0000-0002-7501-3231}}

\address[id=aff1]{Centro de R\'adio Astronomia e Astrof\'isica
  Mackenzie, Escola de Engenharia, Universidade Presbiteriana
  Mackenzie, Rua da Consolac\~ao 896, 01302-907, S\~ao Paulo, Brazil}

\address[id=aff2]{Instituto e Astronom\'ia y F\'isica del Espacio,
  CONICET-UBA, CC. 67 Suc. 28, 1428, Buenos Aires, Argentina}

\address[id=aff3]{SUPA School of Physics \&
  Astronomy, University of Glasgow, Glasgow, G12 8QQ, Scotland}

\runningauthor{J. F. Valle et al.}
\runningtitle{Submillimeter Radiation as the Thermal Component of the Neupert Effect}

\begin{abstract}
  The Neupert effect is the empirical observation that the temporal
  evolution of non-thermal emission (\eg\, hard X-rays) is frequently
  proportional to the time derivative of the thermal emission flux
  (soft X-rays), or vice versa, that time-integrated non-thermal flux
  is proportional to thermal flux. We analyzed the GOES M2.2 event
  \textsf{SOL2011-02-14T17:25}, and we found that the 212\, GHz emission
  plays quite well the role of the thermal component of the Neupert
  effect. We show that the maximum of the hard X-ray flux for energies
  above 50\, keV is coincident in time with the time-derivative of the
  212\, GHz flux, within the uncertainties.  The microwave flux
  density at 15.4\, GHz, produced by optically thin gyrosynchrotron
  mechanism, and hard-X rays above 25\, keV mark the typical impulsive
  phase, and they have similar time evolution. On the other hand, the
  212\, GHz emission is delayed by about 25\, seconds with respect to
  the microwave and hard X-ray peak. We argue that this delay cannot
  be explained by magnetic trapping of non-thermal electrons. With all
  of the observational evidence, we suggest that the 212\, GHz emission
  is produced by thermal bremsstrahlung, initially in the
  chromosphere, and shifting to optically thin emission from the hot
  coronal loops at the end of the gradual phase.
\end{abstract}
\end{opening}

\section{Introduction}

\label{S-Introduction}
The different temporal evolution of hard X-rays (HXR) and soft X-rays
(SXR) during impulsive bursts is known since the \textit{Orbiting
  Solar Observatory}-1 (OSO-1) observations
\citep{White:1964}. Moreover, OSO-1 data also showed that microwaves
(MW) and HXR are time coincident during the impulsive bursts
\citep{Frost:1964}, lending support to their close origin. It was
\citet{Neupert:1968} who noted for the first time that the SXR flux is
better correlated with the time-integrated flux density (fluence) at
the MW frequency $\nu=2.695$\,GHz, \ie

\begin{equation}
F_\mathrm{SXR}(t) \propto \int_{t_\circ}^t F_\mathrm{MW}(t^\prime) \mathrm{d}t^\prime \ ,
\label{eq:neupert}
\end{equation} 
where $F_\mathrm{SXR}$ is the SXR flux and $F_\mathrm{MW}$ is the MW
flux density. This relation holds until $F_\mathrm{SXR}$ reaches its
maximum flux, which should be coincident with the end of the MW
emission.  A similar relation is observed between HXR and SXR
\citep{Hudson:1991}.  This observational fact was interpreted as the
atmospheric response to the heating produced by the energetic
particles when they precipitate into the lower and denser layers:
energetic electrons spiraling within the magnetic fields produce
synchrotron radiation observed at MW, while the HXR is non-thermal
bremsstrahlung produced by Coulomb collisions, transferring energy to
the plasma that expands and emits thermal bremsstrahlung observed at
SXR: a phenomenon also known as chromospheric evaporation
\citep{Neupert:1968,HudsonOhki:1972,Antonuccietal:1984}. Conversely,
Equation \ref{eq:neupert} can be written in terms of the SXR time
derivative \citep{Hudson:1991}:

\begin{equation}
\frac{\mathrm{d}F_\mathrm{SXR}(t)}{\mathrm{d}t} \propto F_\mathrm{MW}(t) \ , \quad \mbox{or} \quad
\frac{\mathrm{d}F_\mathrm{SXR}(t)}{\mathrm{d}t} \propto F_\mathrm{HXR}(t) \ .
\label{eq:neupertD}
\end{equation}

Equations \ref{eq:neupert} and \ref{eq:neupertD} are the
mathematical representation of the Neupert effect. Statistical
analysis of SXR and HXR data show that in around half of the bursts
the effect is present, which means that for around 50\,\% of the cases,
there is evidence for a more complex heating mechanism than only
electron-beam-driven \citep{Veronigetal:2002}.  Furthermore,
\citet{McAteerBloomfield:2013} analyzed the energy band pairs for
which the Neupert effect is better observed, concluding that the best
agreement is between the bands 12--25\,keV (SXR) and 100--300\,keV (HXR).

All of the works mentioned concentrate on the SXR wavelength domain to
observe the thermal emission.  \citet{Trottetetal:2000} observed that
H$\alpha$ has a slow (accumulative) and fast (direct) relationship
with HXR: either chromospheric evaporation or continuous coronal heat
flux may be responsible for the slow response. In another statistical
work, \citet{Veronigetal:2002c} analyzed the timing between SXR, HXR,
and H$\alpha$, showing that in 90\,\% of all cases, SXR starts before
HXR while H$\alpha$ starts simultaneously with HXR. Moreover, SXR and
H$\alpha$ maximum fluxes are temporally coincident with the end of
HXR.

Solar flare observations at submillimeter frequencies (here considered
to be frequencies $> 100$\,GHz) are relatively new. For this reason
the physical origin of the emission is still a subject of
debate. Thermal bremsstrahlung and synchrotron radiation, or a
combination of both, are the main candidate sources of radiation at
high frequencies \citep[see, \eg, ][]{Bastianetal:1998,
  PickVilmer:2008}. Other mechanisms have been suggested
\citep{KaufmannRaulin:2006,FleishmanKontar:2010,Kruckeretal:2013}, but
the lack of a better spectral coverage toward the THz range does not
allow us to draw definitive conclusions. In the work of
\citet{Trottetetal:2002,Trottetetal:2011}, \citet{Luthietal:2004a},
and \citet{Luthietal:2004b} the submillimeter radiation during the
time-extended or gradual phase of the solar flares was compatible with
thermal bremsstrahlung, while during the impulsive phase it was
considered of synchrotron origin. Moreover \citet{Tsapetal:2016} have
shown a particular flare whose emission between 93 and 140\,GHz
increases and can be attributed solely to thermal bremsstrahlung.

We present in this work a peculiar event where submillimeter emission
can be interpreted as the thermal component of the Neupert effect.  We
show that the 212\, GHz time profile is very different from the time
profiles at HXR and MW, that it does not show the typical impulsive
phase \citep{Kruckeretal:2013}, and that its temporal evolution is in
agreement with both Equations \ref{eq:neupert} and
\ref{eq:neupertD}. In contrast to other events
\citep{Luthietal:2004a,Luthietal:2004b,Trottetetal:2011,Trottetetal:2015}
the submillimeter time-extended temporal evolution is not well
correlated with SXR.

\section{Observations and Data Analysis}

\label{sec:observations}

The event \textsf{SOL2011-02-14T17:25} (hereafter
\textsf{SOL2011-02-14} for simplicity) is associated with a GOES
M2.2-class SXR flare in the Active Region 11158. On 14 February, at 0
UT, the region was located at Heliographic Latitude S20 and Longitude
W04, it was 10$^\circ$ wide in longitude, and within it occurred
several C-class events before the M2.2, subject of the present
analysis.  The Solar Submillimeter Telescope
\citep[SST:][]{Kaufmannetal:2008} tracked AR11158 since the day before
the event. On 14 February, the atmospheric conditions were not
favorable for millimeter observations: at 212\, GHz the zenith optical
depth was 0.8, and we can only estimate the lower limit for the
optical depth at 405\,GHz to be $\approx 4.5$. Therefore, at the
antenna elevations during the event, the signal was attenuated 57\,\%
and $> 99$\,\% at 212 and 405\, GHz respectively. In adition to this
correction, the antenna temperatures of the multi-beam array (Beams
2,3 and 4) were subtracted by the antenna temperature of Beam 1, which
is $7^\prime$ far from the array \citep[see ][for a detailed
explanation]{Trottetetal:2011}. While in this way we remove most of
the atmospheric fluctuations, we note that since Beam 1 is off the
active region, a variation in opacity is less amplified by the
\textit{cool} background than the same variation observed by Beams 2,
3 and 4 which are over the \textit{hot} active region. However, at 212\,
GHz active regions are at most 20\,\% hotter than quiet Sun
\citep{Silvaetal:2005}, and therefore this is a limited effect within the
quoted uncertainty.

\begin{figure} 
  \centerline{\includegraphics[width=\textwidth,clip=]{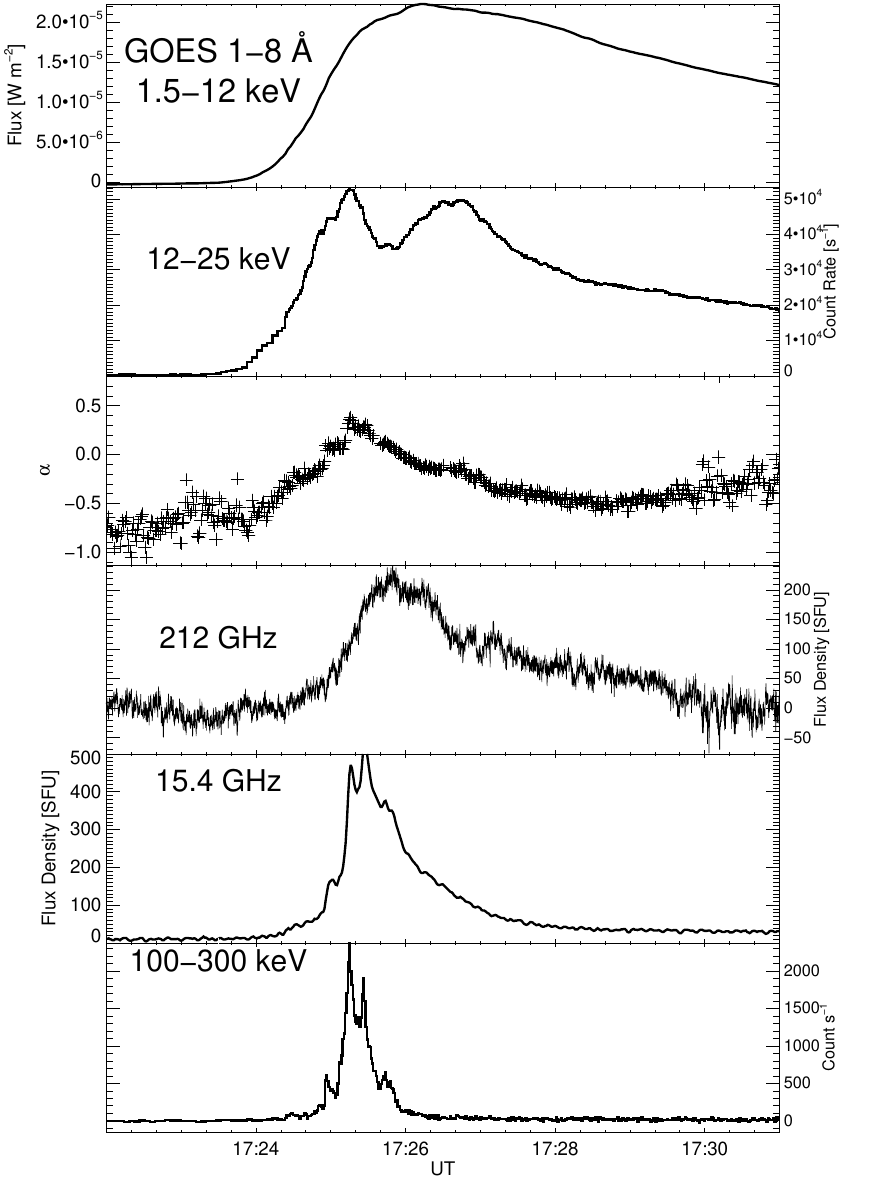}}
  \caption{Temporal evolution at selected frequencies / energies. From
    \textit{top} to \textit{bottom}: GOES 1--8\,\AA\ (1.5--12\,keV);
    \textit{Fermi} GBM NaI 12--25\,keV, spectral index $\alpha$
    obtained between 15.4 and 212\,GHz (see Sec. \ref{sec:spectra}),
    SST 212\,GHz, RSTN 15.4\,GHz, and \textit{Fermi} GBM NaI
    100--300\,keV HXR.  }
  \label{fig:Profiles}
\end{figure}

The SST observed the flare at 212\,GHz with the three beams that
compose its multi-beam system providing the instantaneous emitting
centroid and its flux density every 0.04\,seconds using the technique
described in Section \ref{sec:multibeam}.  The flux density reaches
its maximum between 17:25:45 and 17:25:52~UT (see Figure
\ref{fig:Profiles}), with mean
$\langle Fx_{212}\rangle_\mathrm{max} = 220\pm 30\ \mathrm{SFU}$.  The peak
is relatively smooth, temporal fluctuations around the maximum have a
standard deviation of about 4\,\% of $\langle Fx_{212}\rangle_\mathrm{max}$.

After the peak, the flux density decreases slowly; the event overall
duration is $\approx$ nine minutes. At 405\,GHz no significant signal excess
was detected, a logical consequence of the high optical depth. A rough
estimation of the detectable source flux threshold was obtained from
the measured antenna temperature fluctuations,
$\delta T_{405}\approx 15$\,K, corrected by optical depth and for the
beam offset (that we can infer from the multi-beam solution) and
converted to flux. The result yields $Fx_{405} \gtrsim 10^4$\,SFU. This
number should be considered our uncertainty in the 405\,GHz flux.

The flare has been detected in HXR by the \textit{Gamma-Ray Burst
  Monitor} (GBM) onboard \textit{Fermi} \citep{Meeganetal:2009}. GBM is
composed of twelve sodium-iodide (NaI) and two bismuth-germanium-oxide
(BGO) detectors. In our analysis we used the NaI 128 energy channels
in the range from 4\,keV to 2000\,keV and 1.024\,second time
resolution. Unfortunately there are no \textit{Ramaty High Energy
  Solar Spectroscopic Imager} (RHESSI) data during the impulsive phase
of the event. We complement our analysis with MW from the United
States Air Force (USAF) \textit{Radio Solar Telescope Network}
\citep[RSTN:][]{Guidiceetal:1981} at 1.415, 2.695, 4.995, 8.8, and
15.4\,GHz with one-second temporal resolution, and GOES 1--8\,\AA\
(1.5--12\,keV) with two-second temporal resolution.

\subsection{Temporal Evolution}

Figure \ref{fig:Profiles} presents a selection of the different
frequency / energy temporal profiles. By using GOES and \textit{Fermi}
low energy channels, we cover the whole SXR domain.  At the lowest
energies, GOES 1.5--12\,keV the emission has a smooth evolution
starting at 17:23\,UT, with a simple structure peaking at 17:26 UT,
and returning to pre-flare level not before 18:40\,UT. \textit{Fermi}
12--25\,keV band starts and ends with GOES 1--8\,\AA, but it shows two
peaks: at 17:25:20\,UT and at around 17:26:40\,UT. On the other hand
the impulsive phase at MW and HXR, have a common shape, with delays of
less than one second between the strongest structures. Emission starts
at around 17:24\,UT and ends by 17:27\,UT.

\begin{figure}
\centerline{\includegraphics[width=0.50\textwidth]{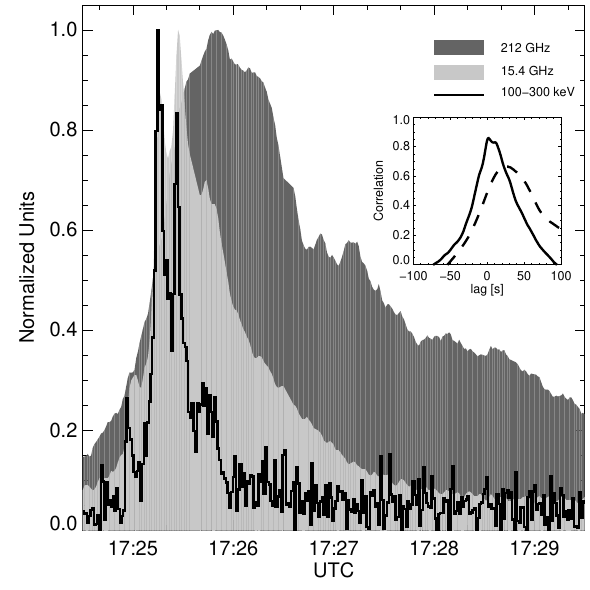}
\includegraphics[width=0.50\textwidth]{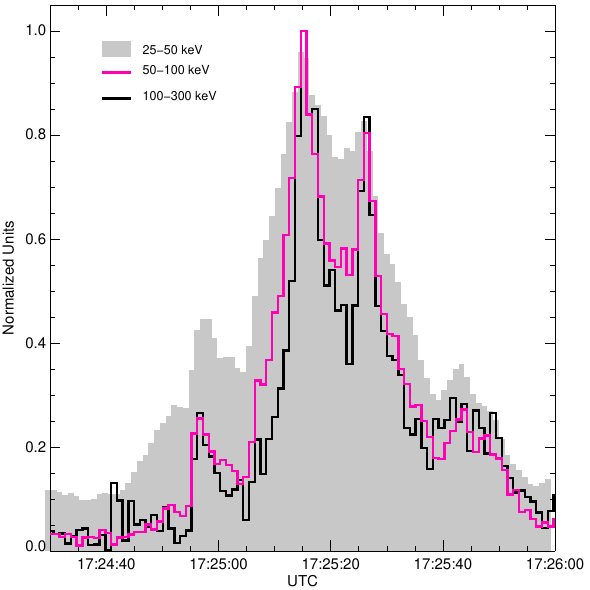}}
\caption{\textit{Left}: Normalized intensities at 212\,GHz (shaded
  dark gray), 15.4\,GHz (shaded light gray) and 100--300\,keV (black
  curve). In the inset, the cross correlation as a function of the lag
  between 100--300\,keV and 15.4\,GHz (continuous curve) and between
  100--300\,keV and 212\,GHz (dashed curve). \textit{Right}:
  Normalized count rates at 25--50\,keV (shaded light gray),
  50--100\,keV (magenta curve) and 100--300\,keV (black curve).}
\label{fig:NProfiles}
\end{figure}

A closer look at the temporal evolution at different frequencies and
energy bands can be seen in Figure \ref{fig:NProfiles}, where the
fluxes are normalized to facilitate the comparison. In the left panel
we present the normalized flux at 212\,GHz (shaded dark gray),
15.4\,GHz (shaded light gray) and 100--300\,keV HXR (black curve). We
observe that during the impulsive phase, defined by HXR, there is an
excellent match between features observed at HXR and 15.4\,GHz, with
no detectable delay.  This implies that 15.4\,GHz has an optical depth
$\tau \lesssim 1$, \ie\ is near the peak of the spectrum.  In
contrast, the submillimeter emission is, as noted before, smoother,
almost featureless, and its peak is delayed by $> 20$\,seconds with
respect to HXR. The cross correlation between HXR and 15.4\,GHz (HXR
and 212\,GHz), graphically exposed in the inset with a continuous line
(dashed line) are a quantitative way to remark the coincidence (lack
of coincidence) of the temporal evolution at different frequencies.
The cross correlation between HXR and 15.4\,GHz is maximum for a
lag=0\,seconds, while between HXR and 212\,GHz is maximum for a
lag=25\,seconds. We have also compared the temporal evolution at the
three HXR energy bands (right panel in Figure
\ref{fig:NProfiles}). The shaded light-gray curve represents the
lowest energy (25--50\,keV), the dark-gray curve the middle energy
(50--100\,keV) and the black curve the highest detected energy
(100--300\,keV). All of the peaks match well; therefore we do not
observe any delay within the data temporal resolution.
 
\subsection{Spectra}
\label{sec:spectra}

\begin{figure}
\centerline{\includegraphics[width=0.50\textwidth]{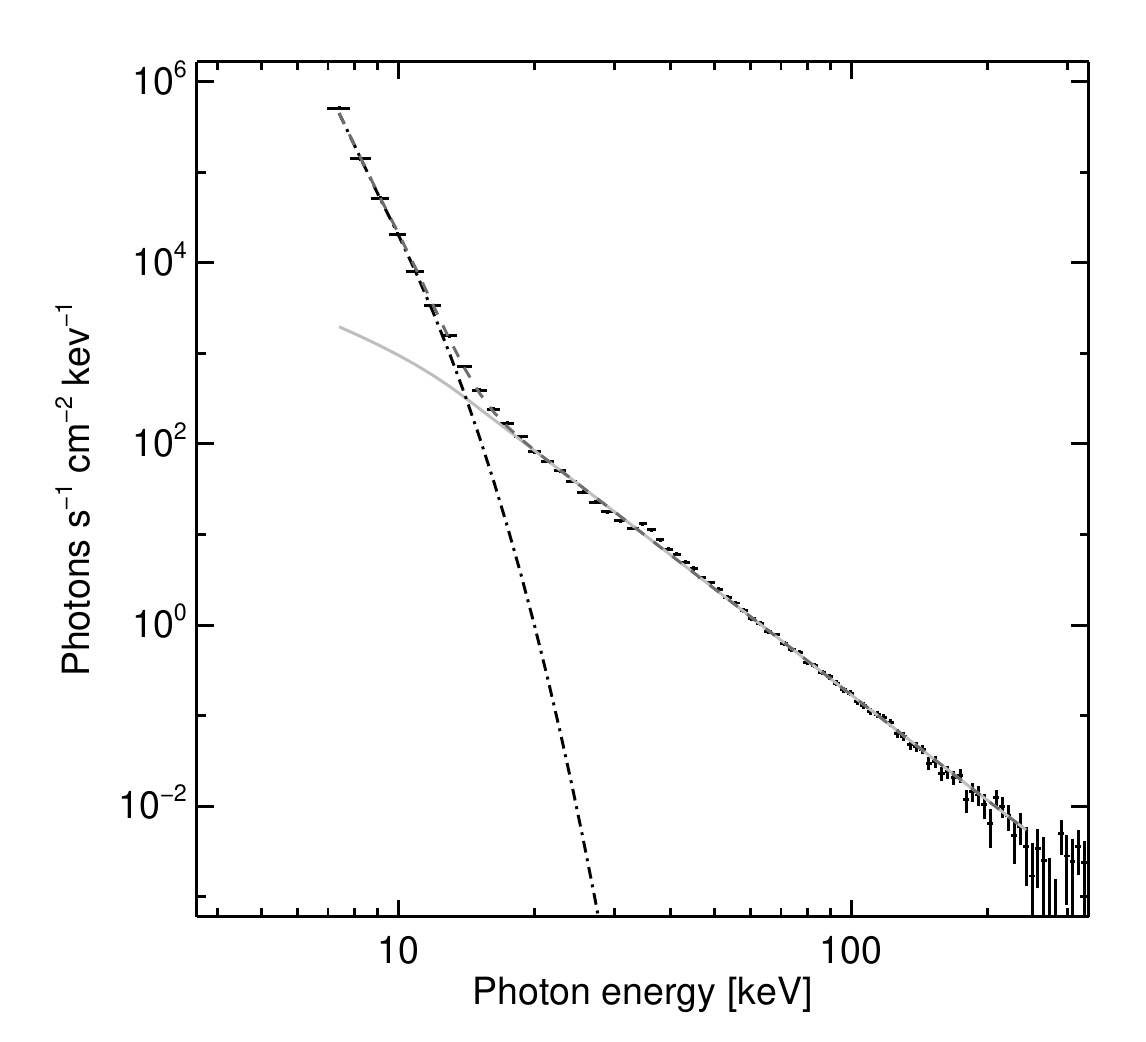}
\includegraphics[width=0.50\textwidth]{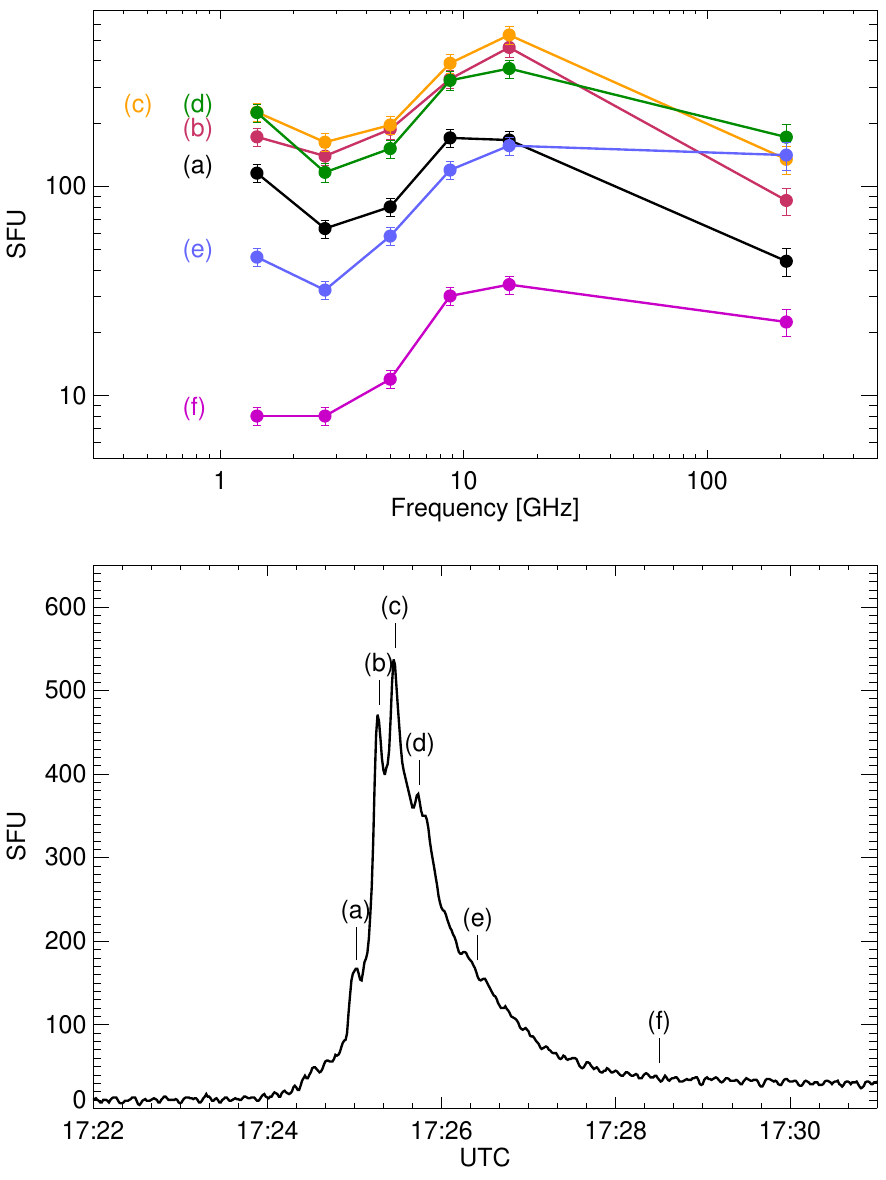}}
\caption{\textit{Left}: HXR photon spectrum at peak time (17:25:24 --
  17:25:28\,UT), the \textit{dashed gray curve} is the fitted photon
  spectrum, \textit{continuous thick gray} represents the single power
  law component, and the \textit{dot--dashed} curve is the thermal
  radiation. \textit{Top right}: Radio spectra at selected one-second
  intervals during the impulsive phase of the flare. \textit{Bottom
    right}: 15.4\,GHz time profile with the selected spectral intervals
  (a to f) identified.}
\label{fig:radio-spectra}
\end{figure}

The \textit{Fermi} GBM HXR fitted photon spectrum during the peak time
four-second interval (17:25:24 -- 17:25:28\,UT) is shown in Figure
\ref{fig:radio-spectra} (left panel). The spectrum is best fitted by a
thermal component (dot--dashed curve) with temperature $T=13$\,MK and
emission measure $\mathrm{EM}=56\times 10^{49}\ \mathrm{cm}^{-3}$,
plus a thick-target component from a power-law distribution of
electrons, with an electron rate of $F=2.6\times 10^{35}$\,s$^{-1}$,
spectral index $\delta = 4.6$; and low-energy cutoff
$E_\mathrm{c}=16$\,keV (continuous gray curve).  From the figure it can
be seen that for energies $>$ 20\,keV the emission can be considered as
non-thermal bremsstrahlung, and that there is no detected emission
above $\approx 200$\,keV.

The radio spectra at different temporal intervals of one-second
duration are shown in Figure \ref{fig:radio-spectra} (right panel). No
firm conclusion can be drawn about the submillimeter emission origin
during the maximum of the microwave emission, intervals b to d,
because of the gap between 15.4 and 212\,GHz. For the same reason we
cannot determine the peak frequency, but, as noted before, the
emission at 15.4\,GHz has an optical depth $\tau_{15.4} \lesssim
1$. After the maximum, intervals e and f, flux density at 8.8,
15.4 and 212\,GHz are very similar, which can be considered an
indication of pure thermal emission. Moreover, while the 15.4\,GHz flux
varies by a factor of around 15 between interval c through f, the
submillimeter emission just halves its flux during the same period. As
an illustration of this behavior we show in Figure \ref{fig:Profiles}
the spectral index between 212 and 15.4\,GHz
$\alpha = - \log{(F_{212}/F_{15.4})}/\log{(212/15.4)}$ as a function of
time. We remark the fact that $|\alpha| \le 0.5$ which is a rather
hard index. Indeed, applying the \cite{Dulk:1985} semi-empirical
formulation for gyrosynchrotron emission, we get an electron index
$\delta \le 1.9$. Moreover, we note that during interval a, before
the peak, $\alpha<0.5$ (and $\delta<2$), indicating again the thermal
origin of the 212\,GHz emission.
    
\subsection{Submillimeter Position}
\label{sec:multibeam}
  
In order to obtain position and flux for our submillimeter
observations we used the iterative multi-beam technique first
introduced by \citet{Herrmannetal:1992} for observations made with the
13.7-m antenna of the \textit{Itapetinga Radio Observatory} and later
applied to SST observations by \cite{Cristianietal:2007}.  The
original method considered point-like sources, and therefore it needed
at least three independent observations.  \citet{Luthietal:2004b}
expanded this method by introducing extended sources, for which they
obtained position, flux, and an effective area using four independent
observations in order to get a unique solution.  In our case we used a
matrix representation of the beams obtained after the deconvolution of
solar maps observed in 2006 following the method developed by
\cite{Costaetal:2002}. We also considered Gaussian extended sources
defined by a four-tuple $(F_\mathrm{s}, x,y,\sigma_\mathrm{s})$ with
$F_\mathrm{s}$ the maximum of the source flux density, $x,y$ its
position, and $\sigma_\mathrm{s}$ its Gaussian standard deviation.

\begin{figure}
  \parbox{0.49\textwidth}{
    \includegraphics[width=0.5\textwidth]{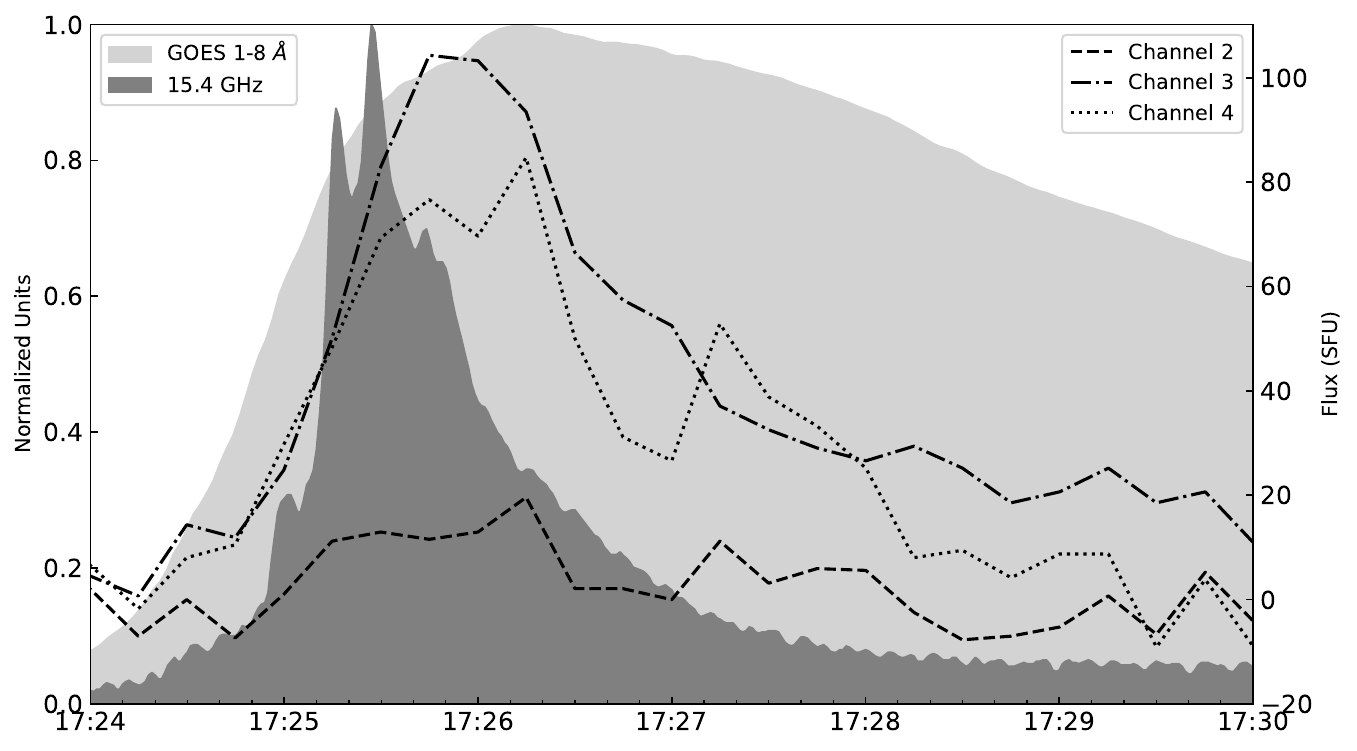}
  }
  \parbox{0.49\textwidth}{
    \includegraphics[width=0.40\textwidth]{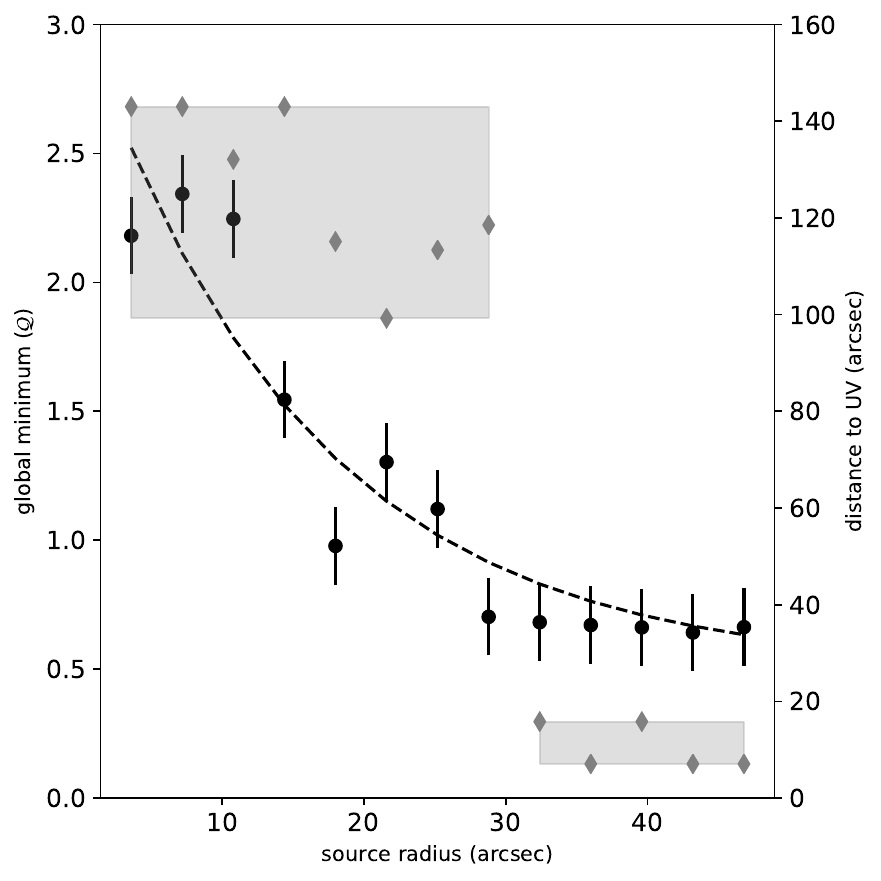}
  }
  \caption{\textit{Left}: Calibrated low-resolution flux density
    temporal profiles used to determine the source flare position. The
    dataset was built picking values every 15\,seconds between
    17:24:00 and 17:30:00\,UT for every receiver of the 212\,GHz
    multi-beam array. Normalized light curves from GOES 1--8\,\AA\ and
    RSTN 15.4\,GHz are shown for comparison. \textit{Right}:
    ${\cal Q}(x_\circ,\circ,\sigma_\mathrm{s})$ normalized to 100 as a
    function of $\varphi=\sigma_\mathrm{s} \sqrt{\ln(4)}$
    (\textit{filled circles}), the \textit{dashed curve} shows the
    global trend. The \textit{diamonds} represent the distance between
    the source positions determined with the multi-beam technique
    $(x_\circ,y_\circ)$ as a function of $\varphi$.}
\label{fig:channels}
\end{figure}

The iterative method compares a combination of measured and
model-calculated relative fluxes of the three different receivers:
\begin{eqnarray}
  Q(t_{i},x,y,\sigma_\mathrm{s}) &=& \sum_{k} \left | \frac{F^\mathrm{meas}_{k}(t_{i})}{F^\mathrm{meas}_{p \neq k}(t_{i})} - \frac{F^\mathrm{cal}_k(t_i,x,y,\sigma_\mathrm{s})}{F^\mathrm{cal}_{p \neq k}(t_i,x,y,\sigma_\mathrm{s})} \right | \ , \\
  F^\mathrm{meas}_{k}(t_{i}) &=& 2\frac{\mathrm{k_B} T_k(t_i)}{A_e} \ , \nonumber
\label{eq:Qmatrix}
\end{eqnarray}
with $T_k(t_i)$ the antenna excess temperature of beam $k$ at instant
$t_i$; $\mathrm{k_B}, \ A_e$ the Boltzmann constant and the antenna
effective area respectively and $F^\mathrm{cal}_k(t_i,x,y,\sigma_\mathrm{s})$ the
corresponding calculated flux obtained after convolving the source
with the beam. We then looked for the location ($x_\circ,y_\circ$)
that overall minimizes $Q$ matrices along the whole event for a fixed
source size. Namely
\begin{equation}
{\cal Q}(x,y,\sigma_\mathrm{s}) = \sum_i Q(t_i,x,y,\sigma_s) \ , \qquad
\left . \frac{\partial^{2}{\cal Q}(x,y,\sigma_\mathrm{s})}{\partial x \partial y}  \right |_{\sigma_s} = 0 \ . 
\label{eq:Qgradient}
\end{equation}
We prepared low-resolution profiles taking values every 15\,seconds
between 17:24:00\,UT and 17:30:00\,UT, as is shown in Figure
\ref{fig:channels}. A grid of
$350^{\prime\prime}\times350^{\prime\prime}$ with $10^{\prime\prime}$
spacing between points is used to calculate $Q(t_i,x,y,\sigma_\mathrm{s})$. The
resulting ${\cal Q}(x_\circ,y_\circ,\sigma_\mathrm{s})$ values,
normalized to 100, as a function of the source angular radius
$\varphi = \sigma_\mathrm{s} \sqrt{\ln(4)}$ are shown in Figure
\ref{fig:channels} (circles). We note that $\cal Q$ decreases when the
source increases in size up to $25^{\prime\prime}$, then stabilizes, \ie\ the method
becomes insensitive to changes in size. The same figure also shows the
distance from $(x_\circ,y_\circ)$ to the center of the biggest and
brightest UV source (diamonds). We observe a similar behavior, when
the source achieves a size of $\approx 30^{\prime\prime}$, the distance
stabilizes around $10^{\prime\prime}$. We conclude that that
$\varphi \ge 25^{\prime\prime}$, which corresponds to a source area
$A_\mathrm{MB} \ge 3.8\times 10^{17} \,\mathrm{cm}^2$. We remark that, since
we do not have four independent observations, we cannot provide a
unique solution such those obtained by \cite{Luthietal:2004b}, or even
by \cite{GimenezdeCastroetal:1999} who worked with analytical
expressions and nominal beams.

\begin{figure}
  \centerline{\includegraphics[width=0.7\textwidth]{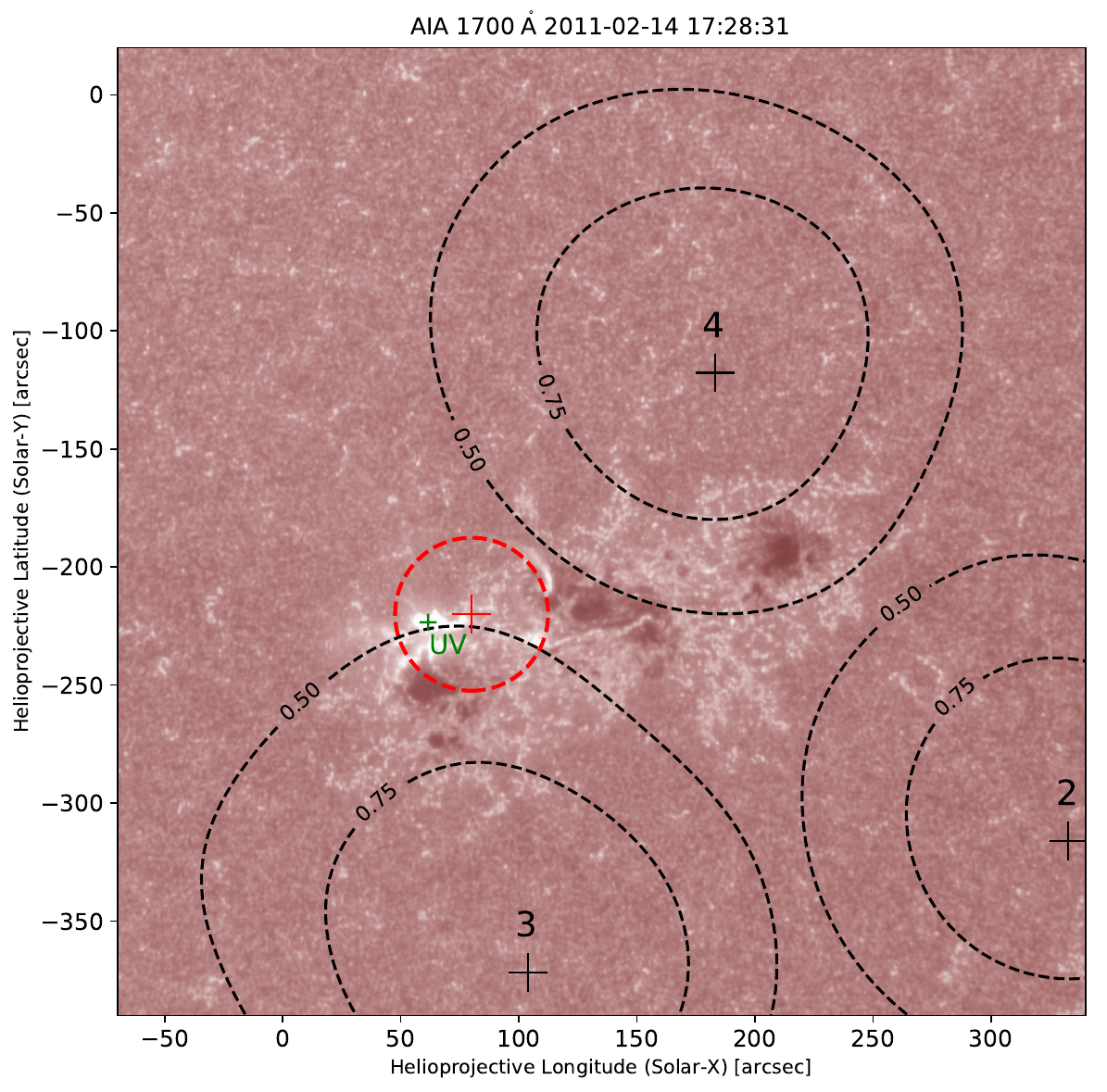}}
  \caption{UV image at 1700\,\AA\ taken by SDO/AIA at 17:28:31\,UT, to
    avoid the saturated pixels. The \textit{red cross} is the solution
    of the iterative multi-beam technique, the \textit{dashed red
      circle} is its absolute uncertainty. The \textit{small green
      cross} shows the center of the brightest and biggest UV
    source. The \textit{dashed black curves} are the three 212\,GHz
    beams used in the multi-beam solution represented at 50\,\% and 75\,\%
    levels; \textit{black crosses} are the beam centers.}
\label{fig:grad-pos}
\end{figure}

In Figure \ref{fig:grad-pos} the submillimeter centroid position for
17:24:00 and 17:30:00\,UT period is shown over a 1700\,\AA\ UV image
as a red cross; the dashed red circle represents its absolute
uncertainty. The small green cross labeled UV marks the centroid
position of the brightest and biggest UV source, and it is the
reference for the distance to the submillimeter solutions.  The excess
flux of the source was obtained after its position and size were
determined.

\section{Origin of the submillimeter emission}

The most common hypothesis for the millimeter and submillimeter
emission during the impulsive phase of solar flares is gyrosynchrotron
from non-thermal electrons. The same electrons should produce HXR,
implying similar HXR and radio light curves; when delays between them
appear, some trapping is considered responsible. However, in this
particular case we do not find evidence of gyrosynchrotron and
trapping in the lightcurves.  Our arguments supporting this conclusion
are:

\begin{enumerate}

\item[\textit{i)}] the temporal evolution at 15.4\,GHz greatly differs from
  that at 212\,GHz, that can be hardly attributed to a $\approx 25$\,s
  electron trapping in the magnetic loop. Had the emission at both
  frequencies been produced by gyrosynchrotron, they would have
  presented similar curves, with or without trapping effects.

\item[\textit{ii)}] There is no indication of $>$ one-second trapping
  when comparing the HXR curves at several energy bands and 15.4\,GHz
  as it is illustrated in Figure \ref{fig:NProfiles}.

\item[\textit{iii)}] The similarities of the high-energy HXR and
  15.4\,GHz curves, \ie\ presence of a number of peaks and lack of
  time delays, indicate that the electrons producing the microwave
  emission are not affected by magnetic trapping. It also suggests
  that the 15.4\,GHz emission is mostly optically thin, and therefore,
  it is, at least, very close to the peak of the gyrosynchrotron
  spectrum. As noted before, the 212\,GHz curve is smooth, and it does
  not follow the 15.4\,GHz.

\item[\textit{iv)}] A 25\,second trapping is a rather extreme
  condition. A long trapping time has a strong impact in other source
  parameters, as the magnetic-field intensity. It has been shown by
  \cite{GimenezdeCastroetal:2009} that the greater the trapping time,
  for a given HXR flux, the smaller the magnetic field. A 25\,s
  trapping would imply a magnetic field of a few Gauss, which would be
  too low to produce any significant emission at 212\,GHz.

\item[\textit{v)}] If the emission had been produced in an homogeneous
  source, the spectral index $\alpha$ between 15.4 and 212\,GHz (Figure
  \ref{fig:Profiles}) would give us information on the electron
  distribution.  However, during the event is always $\alpha < 0.5$,
  which would suggest a very hard, and unlikely, electron distribution
  with $\delta < 1.9$ \citep{Dulk:1985}. Therefore, such a hard
  electron distribution is inconsistent with the value $\delta=4.6$
  derived from the HXR observed spectra. Moreover, a circumstantial
  indication supporting our conclusion is the absence of observational
  evidence for $> 1$\,MeV electrons in the HXR data: the highest HXR
  energy detected  significantly is $\approx 200$\,keV. It is well known
  that relativistic electrons are the main source of the 212\,GHz
  synchrotron \citep[\eg,][]{Ramatyetal:1994, Trottetetal:2015}.

\end{enumerate}

\subsection{The Impulsive Phase}

Taking into account all of the arguments presented above, we conclude
that it is very unlikely that 212\,GHz emission is due to
gyrosynchrotron. This leaves thermal brems\-strahlung as the most
likely mechanism to produce the observed 212\,GHz radiation. The
emission at these frequencies, during the gradual phase of flares, has
been interpreted as due to thermal brems\-strahlung
\citep{Trottetetal:2002,Trottetetal:2011,Luthietal:2004a,Luthietal:2004b,Tsapetal:2016}. Our
observations and analysis put thermal bremsstrahlung as the dominating
mechanism during the impulsive phase of \textsf{SOL2011-02-14} as
well.

It has been shown that ionized plasmas at $T<1$\,MK in the
chromosphere are efficient mm-wave sources
\citep{Kasparovaetal:2009,HeinzelAvrett:2012, Simoesetal:2017}. In
particular, \cite{Simoesetal:2017} have shown that once the energy
deposition in the chromosphere stops, the free electrons quickly
recombine with ions, thus reducing the main source of free--free
emission after the impulsive phase. This decrease of the chromospheric
emission could allow the optically thin free--free emission from the
coronal plasma to dominate late in the gradual phase. and produce the
microwave spectrum observed at the time intervals e and f in Figure
\ref{fig:radio-spectra}.

Moreover, \cite{Trottetetal:2015} interpreted that most of the 30\,THz
radiation observed during the flare \textsf{SOL2012-03-13} is thermal
bremsstrahlung of an optically thin source located above the
temperature minimum in a $T\approx 8000$\,K plasma heated by the energy
deposited by precipitated particles (electrons, protons and
$\alpha$). \cite{Simoesetal:2017} have reached similar conclusions,
using their results from numerical modeling to interpret the
mid-infrared flare reported by \cite{Pennetal:2016}. Extending their
calculations into the sub-THz range, they show that, during the
impulsive phase, the sub-THz emission would be associated with the
upper chromosphere with temperature around $T \approx
10^{4.6}$~K. Therefore, sub-THz as the thermal
counterpart of the Neupert effect is certainly possible.

We present simple calculations to show that the observed flux density
at 212~GHz can be easily explained by optically thick free--free
emission originating in the upper chromosphere. The observed flare
excess $\Delta S_\mathrm{f}$ is simply the difference between the total flux
during the flare $S_\mathrm{f}$ and pre-flare $S_\mathrm{b}$, as observed by the SST
212\,GHz beam:

\begin{equation}
\Delta S_\mathrm{f} = S_\mathrm{f} - S_\mathrm{b} \ . 
\end{equation}
Using the Rayleigh--Jeans law, this becomes: 
\begin{equation}
\Delta S_f = \frac{2\mathrm{k_b} \nu^2}{\mathrm{c}^2}\left(\frac{T_\mathrm{f} A}{D^2}-T_\odot\Omega_\mathrm{b} \right)
\label{eq:RJ}
\end{equation}
where c is the speed of light, $T_\mathrm{\odot}$ is the brightness temperature
of the quiet Sun at $\nu=212$\,GHz, $\Omega_f=A/D^2$ is the solid angle
of the flare, for an area $A$ and
$\Omega_\mathrm{b}=1.06 \ 10^{-6} \ \mathrm{str}$, is the solid angle of the
$4^\prime$ beam angular diameter, and finally $D$ is the Sun--Earth distance,
one astronomical unit (AU). With numerical values,
$\Delta S_\mathrm{f} = 220$\,sfu, $T_\mathrm{s}=5,500$\,K (Figure 3 of
\opencite{Selhorstetal:2005}), and rearranging, Equation~\ref{eq:RJ}
becomes:
\begin{equation}
A_{20} = \frac{1.66}{T_4},
\label{eq:simple}
\end{equation}
where $A_{20}$ is the flare area in $10^{20}$\,cm$^{2}$ and $T_4$ is
the brightness temperature at 212\,GHz in $10^4$\,K.

Following the procedure introduced by \cite{Simoesetal:2017}, we
calculated the brightness temperature $T_\mathrm{b}$ at 212\,GHz for
two models of the \textsf{F-CHROMA} flare model database
(\url{www.fchroma.org/?page_id=24}). The database contains more than
90 flare models and it was generated using the code \textsf{RADYN}
\citep{CarlssonStein:1995, Allredetal:2015}, starting from a quiet Sun
atmospheric model based on VAL-C \citep{VAL81}.  \textsf{RADYN} solves
the coupled, non-linear, equations of hydrodynamics, atomic level
populations, radiative transfer in a 1D atmosphere subject to energy
input by a beam of accelerated electrons. The electron transport and
energy deposition is treated by solving the Fokker--Planck equation
for an initial electron power--law distribution with spectral index
$\delta$, low-energy cutoff $E_\mathrm{c}$, with an energy flux
$F$. We only present a brief description of the \textsf{RADYN} code
here and note that \cite{Allredetal:2015} should be consulted for more
details.

We selected the models with electron-beam parameters closer to the
ones estimated from the HXR spectral analysis, namely, Model 21
($\delta=5$, $E_\mathrm{c}=15$\,keV, $F=10^{10}$\,erg
s$^{-1}$\,cm$^{-2}$), and Model 27 ($\delta=5$, $E_\mathrm{c}=15$\,keV,
$F=3\times 10^{10}$\,erg s$^{-1}$\,cm$^{-2}$). The maximum $T_\mathrm{b}$
values found are $\approx 6.7\times10^4$\,K and
$\approx 10\times10^4$\,K, for Models 21 and 27, respectively. Using
these values for $T_4$ in Equation~\ref{eq:simple} gives an estimate of the
necessary emitting areas to produced the observed flare emission of
220\,sfu: $A=2.5\times10^{19}$ and $A=1.6\times10^{19}$\,cm$^{2}$, for
Models 21 and 27, respectively.

\begin{figure}
  \centerline{\includegraphics[width=\textwidth]{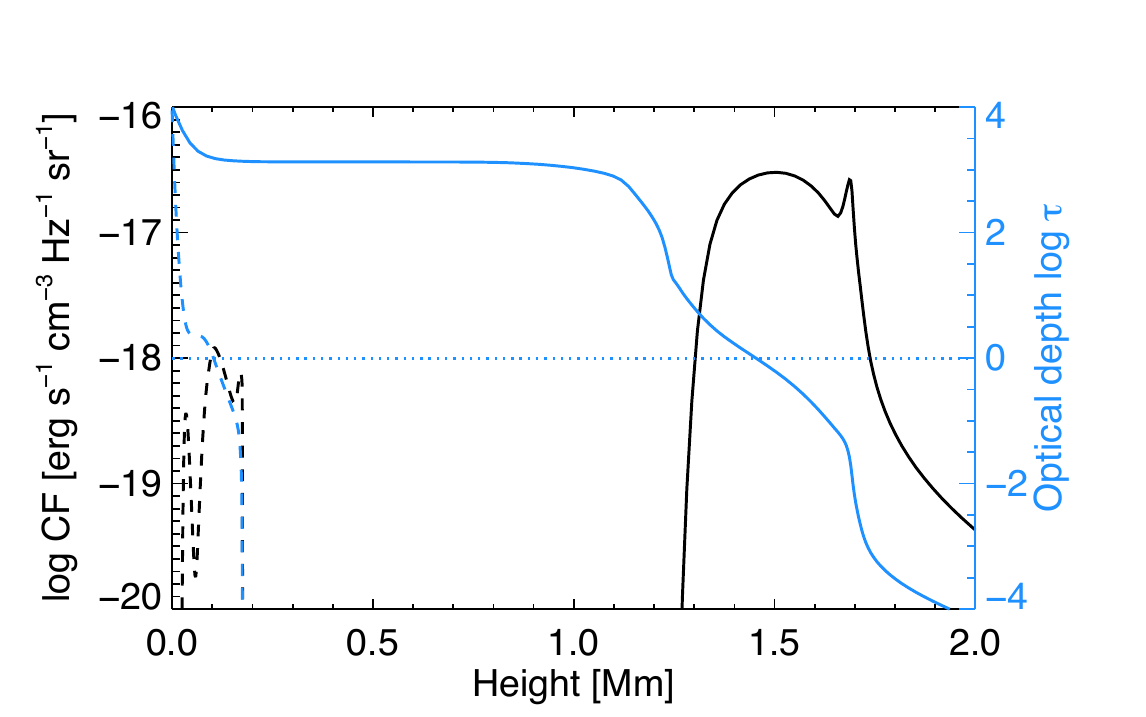}}
  \caption{Contribution function [CF] (\textit{black}) and optical depth
    $\tau$ (\textit{blue}) at 212\,GHz calculated from the
    \textsf{F-CHROMA} flare model 21, at the time of maximum energy
    input ($t=10$\,seconds, \textit{continuous lines}) and pre-flare
    (\textit{dashed lines}) for reference. The \textit{dotted}
    horizontal line shows $\tau=1$.}
\label{fig:cf}
\end{figure}

The resulting contribution function [CF] and optical depth $[\tau]$ for
212\,GHz are shown in Figure~\ref{fig:cf}. [CF] indicates the formation
height of the radiation \citep[\eg,][]{Carlsson:1998, Simoesetal:2017},
which originates near the base of the photosphere in the quiet Sun but
shifts to the upper chromosphere during a flare. The emission is
optically thick in both cases as indicated by the large optical depth
$\tau>1$.

The area of the flaring chromosphere can be estimated from AIA
images. The best \textit{Atmospheric Imaging Assembly} (AIA) band for
this purpose is 1700\,\AA, since, unfortunately, most AIA bands
saturate heavily during this event, especially during the impulsive
phase. According to \cite{Simoesetal:2019}, the flare-excess emission
captured by the AIA 1700\,\AA\ band originates in the chromosphere. To
estimate the flaring area, we constructed histograms of AIA 1700\,\AA\
images and subtracted the average histogram of pre-flare images from
the flare histograms. This resulted in the total number of pixels with
values enhanced by the flare. The flaring area is then simply obtained
by adding all the histogram bins and multiplying the result by the
area relative to the AIA pixel ($0.6 \times 0.6$ arcsec$^2$
corresponding to $\approx 1.9\times 10^{15}$\,cm$^2$). Ignoring the
saturated images, we found an average area of
$A \approx 2.6 \times 10^{19}$~cm$^2$, which is sufficient to produce
the maximum observed flux density ($\approx 220$~sfu), with the
assumed $T_4$ values above; and also much bigger than the lower limit
value $A_\mathrm{MB}$ obtained from the multi-beam technique.

These calculations are not an attempt to model this specific event. Our
goal is show that typical flare characteristics are sufficient to
generate an observable flare signature at sub-mm from optically thick
free--free from the upper chromosphere.

\begin{figure}
\centerline{\includegraphics[width=\textwidth]{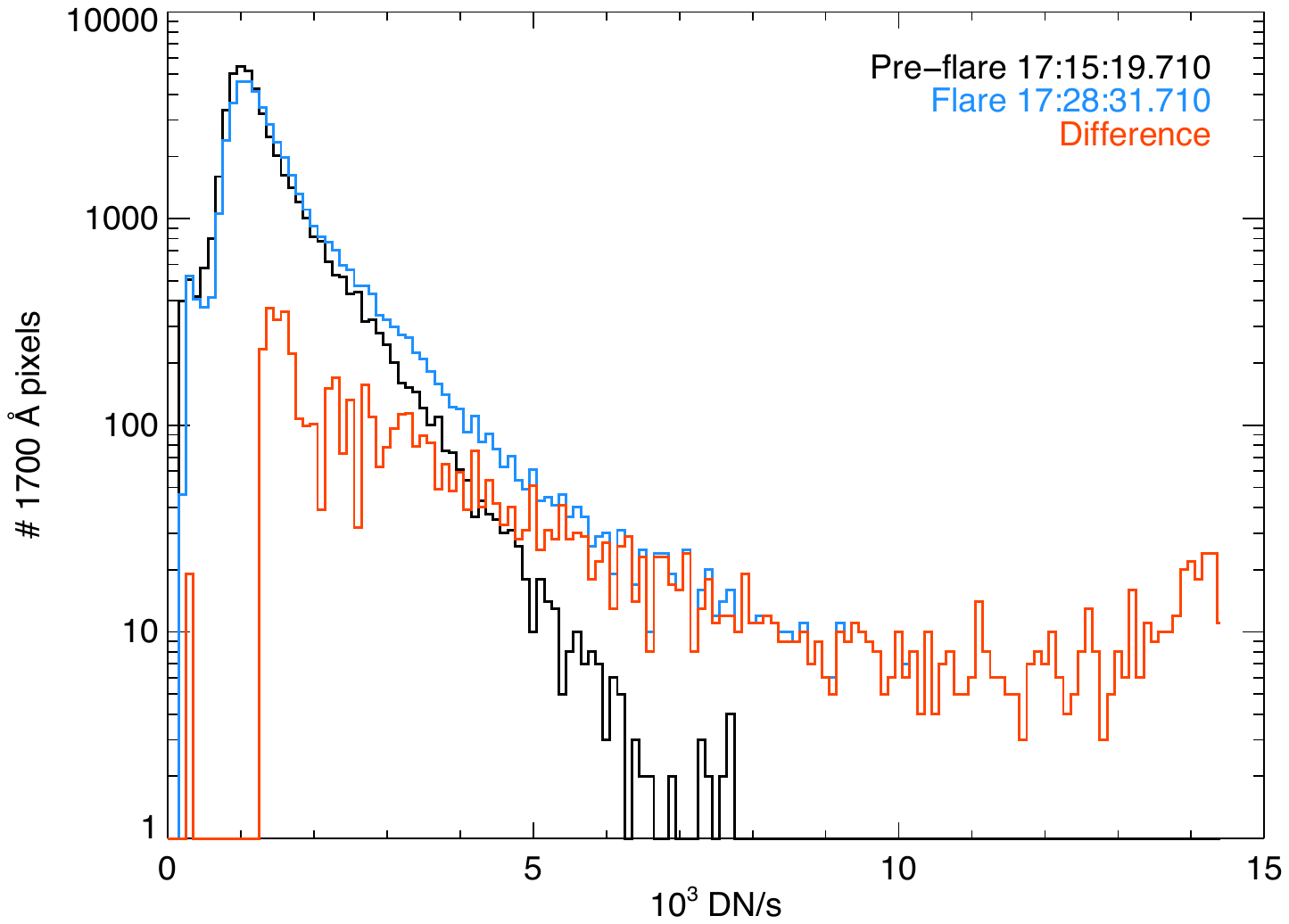}}
\caption{Histograms of pixel values $[\mathrm{DN\,s^{-1}}]$ of AIA
  1700\,\AA\ images at pre-flare (\textit{black}) and flare times
  (\textit{blue}). We used their difference (\textit{orange}) to estimate the
  number of flaring pixels and hence the flaring area.}
\label{fig:area}
\end{figure}

\subsection{The Gradual Phase}

After the end of the HXR emission (around 17:26), the radio spectrum
from microwaves to the submillimeter is consistent with thermal
bremsstrahlung. Even if we lack intermediate frequencies between 15.4
and 212\,GHz to better characterize the spectrum, it is likely that
for $\nu \ge 15$\,GHz the emission is optically thin with a flux
$F_\mathrm{gradual} \approx 40-50$\,SFU. From GOES data we obtain
$EM_\mathrm{goes} \sim 10^{49}\, \mathrm{cm}^{-3}$ and
$T_\mathrm{goes} \approx 16$\,MK. If the radio source were coronal, it
would produce a flux density at 212\,GHz which is a tenth of the
observed one. This problem was observed already in other works
\citep[\eg,][]{Luthietal:2004a,Cristianietal:2007}. \cite{Trottetetal:2011}
addressed the same question for \textsf{SOL2003-10-27T12:30} and
solved the problem using a multi-thermal coronal source, where the
lower layers are cooler. In this way the relatively high submillimeter
density flux was explained. We note that \textsf{SOL2003-10-27T12:30},
an M6.7 GOES class event, has a gradual phase with a similar density
flux at submillimeter frequencies as \textsf{SOL2011-02-14}, and that
the GOES emission measure ($\approx 10^{49} \,\mathrm{cm}^{-3}$) and
temperature ($\approx 12$\,MK) are also quite similar. Therefore,
conclusions from \cite{Trottetetal:2011} can also be applied to this
work.

\subsection{Neupert Effect}

\label{sec:neupert}

\begin{figure*}
\centerline{\includegraphics[width=0.49\textwidth]{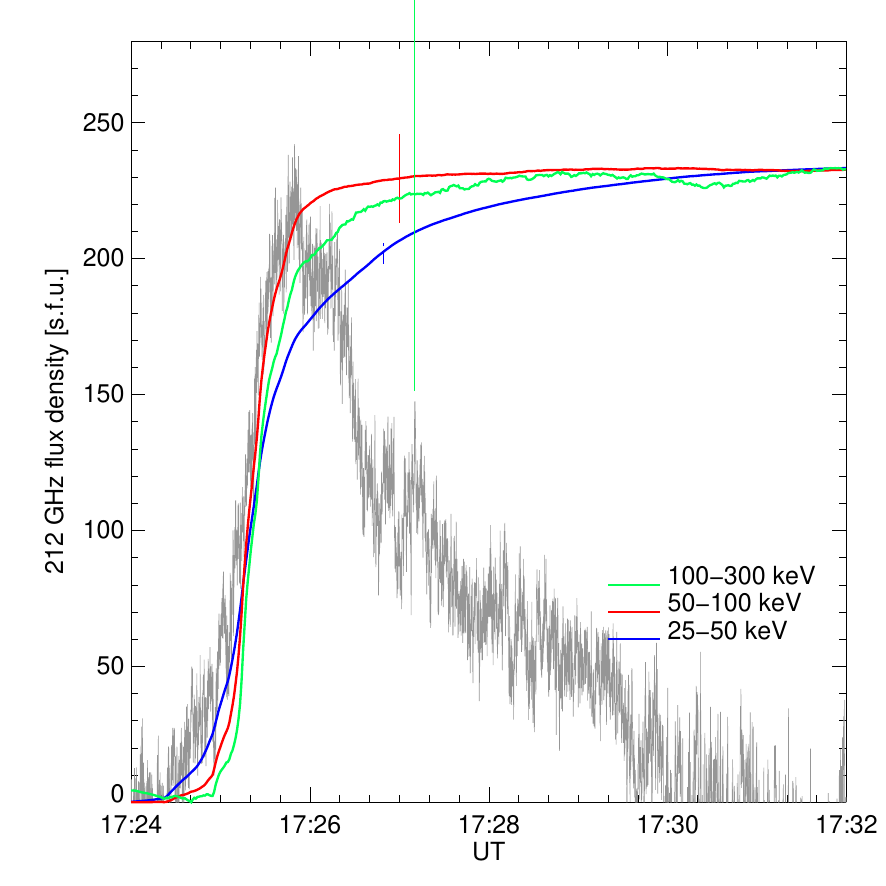}
\includegraphics[width=0.49\textwidth]{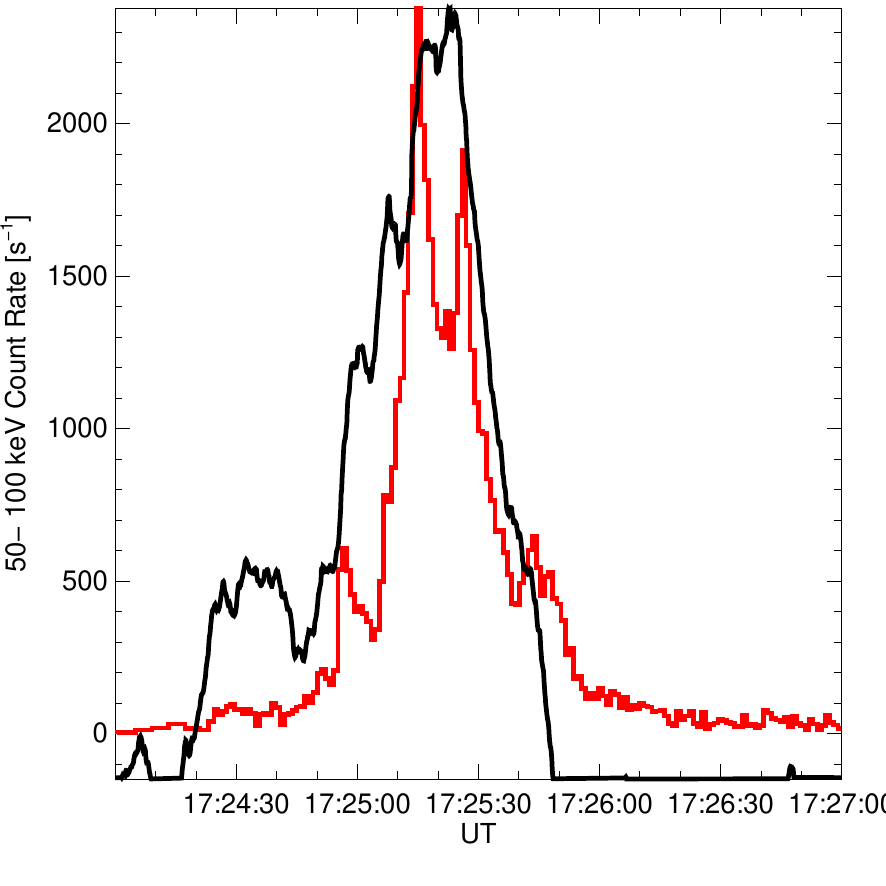}}
\centerline{\includegraphics[width=0.49\textwidth]{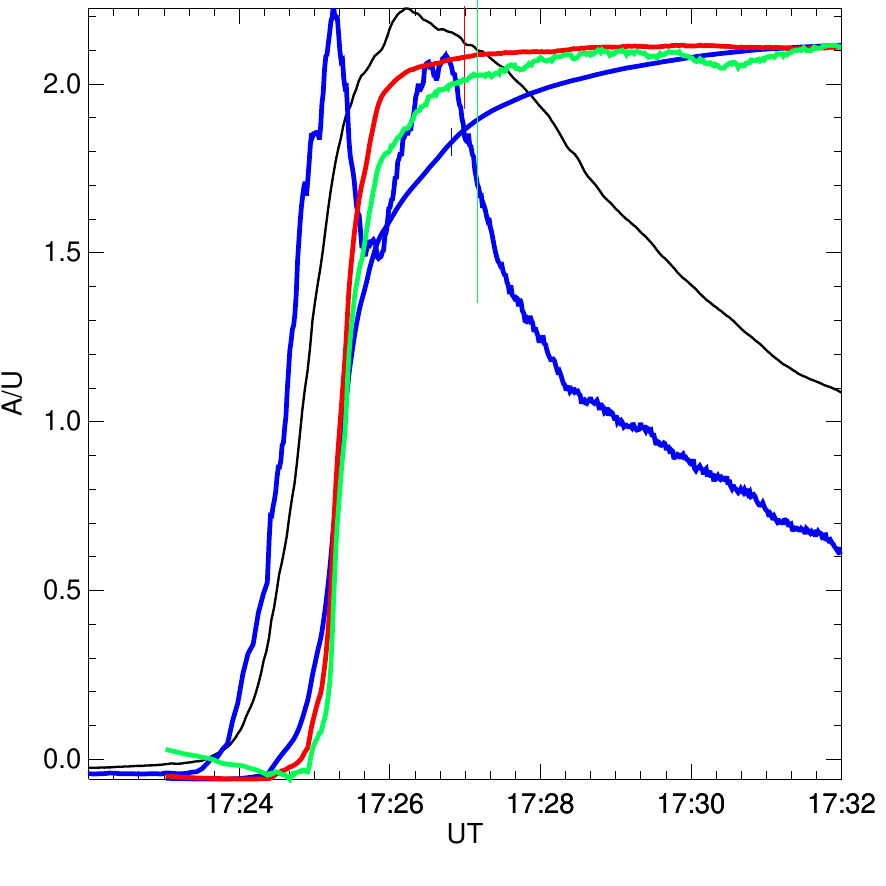}
\includegraphics[width=0.49\textwidth]{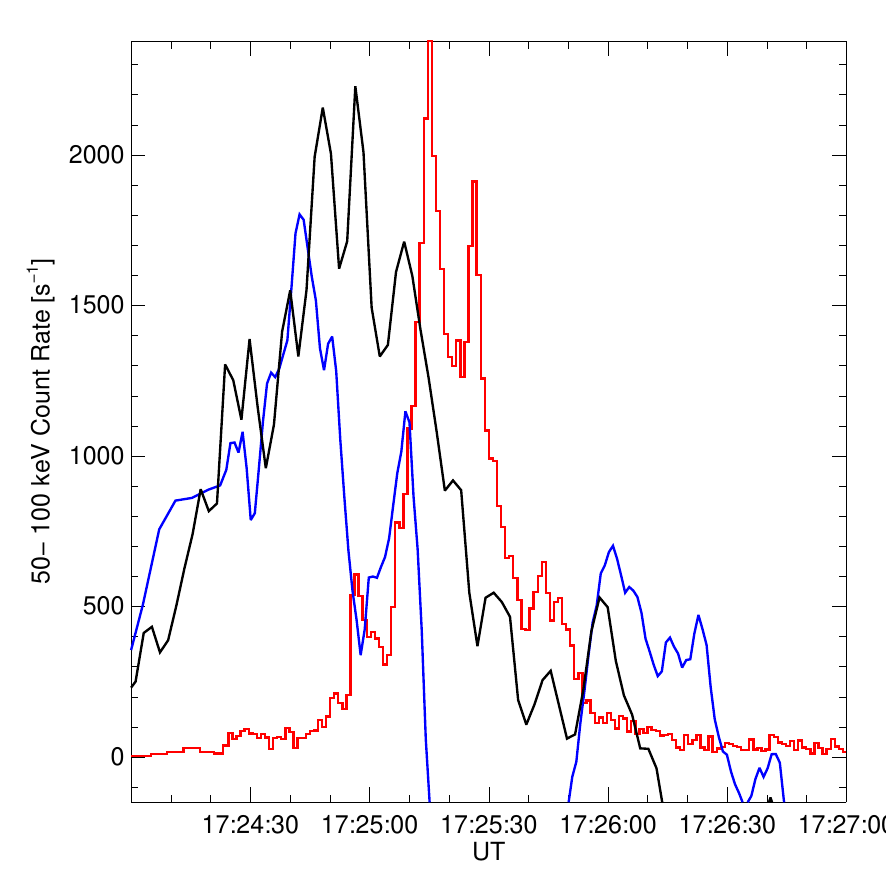}}
\caption{\textit{Top-left}: The flux density at 212\,GHz
  (\textit{gray}) plotted along with the time integrated HXR count
  rate at three different energy bands: 25--50\,keV (\textit{blue});
  50--100\,keV (\textit{red}), and 100--300\,keV
  (\textit{green}). Vertical bars, with the same colors as the curves,
  represent their instantaneous uncertainty. \textit{Top-right}:
  50--100\,keV count rate (\textit{red}) plotted along with the
  12\,seconds smoothed 212\,GHz flux-density tempoarl derivative
  (\textit{black}) positive values. \textit{Bottom-left}: the
  1.5--12\,keV GOES flux (\textit{black}), plotted along with the
  12--25\,keV \textit{Fermi} count-rate (\textit{blue}), and the
  time-integrated HXR count-rate at the same three energy bands of the
  top panel.  \textit{Bottom-right}: 50--100\,keV count rate
  (\textit{red}) plotted along with the SXR temporal derivatives;
  \textit{blue} is for GOES 1.5--12\,keV, and \textit{black} is for
  \textit{Fermi} 9--12\,keV count rate.}
\label{fig:Neupert}
\end{figure*}

We binned the \textit{Fermi} NaI 128 energy channels in three HXR
bands: 25--50\,keV, 50--100\,keV, and 100--300\,keV. These bands were
chosen in order to avoid the contribution from the thermal emission
below 20\,keV and the noise above 300\,keV. We numerically integrated
the HXR counts between $t_0=$~17:23\,UT and 17:32\,UT. The resulting
curves were compared with the 212\,GHz flux-density temporal evolution
(Figure \ref{fig:Neupert}, top left) and with the SXR GOES
1.5--12\,keV and Fermi 12--25\,KeV (Figure \ref{fig:Neupert}, bottom
left). We note a coincidence between the maximum of the fluence
50--100\,keV and the peak at 212\,GHz. For the 25--50\,keV band the
fluence only reaches its maximum at the end of the time
interval. Nonetheless it is worth to note that it follows the initial
212\,GHz emission curve between 17:24\,UT and 17:25\,UT remarkably
well. The association with SXR, however, is not so good. We observe
that the SXR emission starts about 30 s before the HXR fluence.  GOES
1.5--12\,keV peak occurs simultaneously with the HXR fluence,
meanwhile the \textit{Fermi} 12--25\,keV first peak is 30\,seconds behind, only the
second less intense peak is coincident with the fluence maximum.

We computed the time derivative (Equation \ref{eq:neupertD}) of the
212\,GHz flux density using a three-point quadratic Lagrangian
interpolation for unevenly spaced data and plotted its positive values
along with the 50--100\,keV count rate (Figure \ref{fig:Neupert}, top
right panel). We have smoothed the 212\,GHz data, using a 12\,seconds running
mean, to reduce the noise from the derivative procedure. Applying the
same procedure to the SXR reveals a different behavior (Figure
\ref{fig:Neupert} bottom-right).

\section{Conclusions}

We present in this work a peculiar event observed at 212\,GHz whose
emission can be attributed to thermal bremsstrahlung during the
impulsive and gradual phases. During the impulsive phase the 212\,GHz
emission comes from a thermal source at chromospheric heights.
Moreover, its temporal derivative mimics the HXR flux (Figure
\ref{fig:Neupert}, top-right), conforming with the thermal counterpart
of the Neupert effect. During the extended phase, the sub-THz emission
might be characterized as optically thin thermal
bremsstrahlung, from a coronal multi-thermal source as described by
\cite{Trottetetal:2011}.

Differently from the sub-THz, in the SXR domain, we observe that
during the event the emission starts well in advance of the HXR energy
accumulation (Figure \ref{fig:Neupert} bottom-left). Moreover, its
temporal derivatives are significantly different from the HXR temporal
evolution.  This is not completely unexpected, since statistics show
that 50\,\% of all events observed at SXR do not follow the Neupert
effect hypothesis \citep{Veronigetal:2002}. In our case, SXR starts
before the HXR fluence, which can be interpreted as a pre-heating of
the plasma \citep{Veronigetal:2005}. That the temporal evolution of the
sub-THz bremsstrahlung and SXR are not always coincident during a
flare is already known
\citep[\eg\ ][]{Trottetetal:2002,Tsapetal:2016}, which explains why the
Neupert effect is observed at 212\,GHz and not at SXR.

We note that this is the first time that the iterative multi-beam
technique has been used to deduce position, flux, and effective area
with the SST.  \citet{Cristianietal:2007}, used the SST iterative
method for point like sources, and the derived positions were
compatible with the magnetic structures that originated the flare. In
the present case, we found that in order to minimize the difference
between expected and observed fluxes, the matrices
$Q(t_i,x,y,\sigma_\mathrm{s})$, we need an extended source with
$\varphi \ge 25^{\prime\prime}$. At the same time, and reinforcing the
result, this solution is the closest to the main UV emitter (Figures
\ref{fig:channels} and \ref{fig:grad-pos}). As we pointed out above,
we don't find a unique solution here because we do not have four
independent observations; we got a lower limit bound instead. The fact
that this solution stabilizes above a threshold
$\varphi\ge 25^{\prime\prime}$ is an indication of an extended source
\citep{GimenezdeCastroetal:1999}.
  
As a final remark, we stress the importance of the submillimeter and
THz observing range for energy transport diagnostics in the solar
atmosphere.  Observations with the new instrumentation, at
submillimeter frequencies \citep[\eg\ ALMA:][]{Wedemeyeretal:2016},
THz/infrared
\citep{Kaufmannetal:2013,Kaufmannetal:2016,Pennetal:2016}, and/or
near-infrared \citep{Kleintetal:2016} should help the theoretical
works to produce more refined flare models helping to increase our
knowledge about the flaring solar atmosphere.


\begin{acknowledgments}

  The authors are grateful to Hugh Hudson for his enlightening
  comments about the Neupert effect and its history.  J.F. Valle Silva
  acknowledges FAPESP for the support during his PhD Thesis (grant
  2012/1619-9) and CAPES for the Postdoctoral PNPD
  contract. G. Gim\'enez de Castro and J.-P. Raulin acknowledge CNPq
  (contracts 305203/2016-9 and 312066/2016-3).  The research leading
  to these results has received funding from the European Community's
  Seventh Framework Program (FP7/2007-2013) under grant agreement
  no. 606862 (\textsf{F-CHROMA}), CAPES grant 88881.310386/2018-01, FAPESP grant 2013/24155-3 and the
  US Air Force Office for Scientific Research (AFOSR) grant
  FA9550-16-1-0072. AIA data is courtesy of NASA/SDO and the AIA, EVE,
  and HMI science teams. This work is based on data acquired at
  Complejo Astron\'omico El Leoncito, operated under agreement between
  the Consejo Nacional de Investigaciones Cient\'\i ficas y T\'ecnicas
  de la Rep\'ublica Argentina and the National Universities of La
  Plata, C\'ordoba and San Ju\'an.

\end{acknowledgments}

\section*{Disclosure of Potential Conflicts of Interest}
The authors declare that they have no conflicts of interest.

\bibliographystyle{spr-mp-sola}

\end{article} 

\end{document}